\documentclass[12pt]{article}
\usepackage{times}
\usepackage[numbers,sort]{natbib}
\usepackage{epsfig, amssymb, amsmath, multicol}
\usepackage{color}
\usepackage{wrapfig} 
\usepackage{hyperref} 

\newcommand{\citepeg}[1]{\citep[{e.g.,}][]{#1}}

\long\def\@makecaption#1#2{\vskip 2ex\noindent#1\ #2\par}% 
\newcommand\figcaption{\@testopt{\@xfigcaption}{}}% 
\def\@xfigcaption[#1]#2{{\def\@captype{figure}\caption{#2}}}% 

\usepackage{geometry}
\usepackage[layout=modern]{advancedcoverpage}

\newcommand{\fb}[1]{
  \begin{center}
    \fbox{\parbox{.95\textwidth}{#1}}
  \end{center}
}
\title{Coordinated Science in the Gravitational and Electromagnetic Skies}
\subtitle{A Whitepaper Submitted to the Decadal Survey Committee}
\presentedto{\underline{Science Frontier Panels:} \\
{\it PRIMARY: Cosmology and Fundamental Physics (CFP)} \\
{\it SECONDARY: Stars and Stellar Evolution (SSE)} \\ 
{\it ~~~~~~~~~  Galaxies across Cosmic Time (GCT)}}

\author{\emph{\underline{Authors}}\\
Joshua S.\ Bloom, Department of Astronomy, UC Berkeley\\
Daniel E.\ Holz, Theoretical Division, Los Alamos National Laboratory\\
Scott A.\ Hughes, Department of Physics, MIT\\
Kristen Menou, Department of Astronomy, Columbia University,\\ ~~ \\
Allan Adams (MIT), Scott F. Anderson (Univ. of Washington), Andy Becker
(U. Washington), Geoffrey C. Bower (UC Berkeley), Niel Brandt (Penn State),
Bethany Cobb (UC Berkeley), Kem Cook (Lawrence Livermore National
Laboratory/IGPP), Alessandra  Corsi (INAF-Roma), Stefano Covino
(INAF-Osservatorio Astronomico di Brera), Derek Fox (Penn State University),
Andrew Fruchter (STSCI), Chris Fryer (Los Alamos National Laboratory), Jonathan
Grindlay (Harvard/CfA), Dieter Hartmann (Clemson), Zoltan Haiman (Columbia),
Bence Kocsis (IAS), Lynne Jones (U. Washington), Abraham Loeb (Harvard),
Szabolcs Marka (Columbia University), Brian Metzger (UC Berkeley), Ehud Nakar
(Tel Aviv University), Samaya Nissanke (CITA, Toronto), Daniel A. Perley (UC
Berkeley), Tsvi Piran (The Hebrew University), Dovi Poznanski (UC Berkeley), Tom
Prince (Caltech), Jeremy Schnittman (JHU), Alicia Soderberg (Harvard/CfA),
Michael Strauss (Princeton), Peter S. Shawhan (University of Maryland), David
H. Shoemaker (LIGO-MIT), Jonathan Sievers (CITA, Toronto), Christopher Stubbs (Harvard/CfA), Gianpiero Tagliaferri (INAF-Osservatorio Astronomico di Brera), Pietro Ubertini (INAF-Roma), and Przemyslaw Wozniak (Los Alamos National Laboratory)}

\organization{\underline{Projects/Programs Emphasized:} \\
1. The Energetic X-ray Imaging Survey Telescope (EXIST);
\href{http://exist.gsfc.nasa.gov}{\tt http://exist.gsfc.nasa.gov} \\
2. The Synoptic All-Sky Infrared Imaging Survey (SASIR);
\href{http://sasir.org}{\tt http://sasir.org} \\
3. The Large Synoptic Survey Telescope (LSST); \href{http://lsst.org}{\tt http://lsst.org}\\
4. The Laser Interferometer Space Antenna ({\it LISA});
\href{http://lisa.nasa.gov}{\tt http://lisa.nasa.gov}}
\date{}

\begin{document}
\maketitle

\def\ale{\mathrel{\hbox{\rlap{\hbox{\lower4pt\hbox{$\sim$}}}\hbox{$<$}}}}
\def\age{\mathrel{\hbox{\rlap{\hbox{\lower4pt\hbox{$\sim$}}}\hbox{$>$}}}}

\newcommand\ion[2]{#1$\;${\small\rmfamily\sc{#2}}\relax}% 

\newcommand{\avgchi}{\bar{\chi}_{\rm H I}}
\newcommand{\barxi}{{\bar{x}_i}}
\newcommand{\Mpc}{{\rm Mpc}}
\newcommand{\MHz}{{\rm MHz}}
\newcommand{\Msun}{{M_{\odot}}}
\newcommand{\bfM}{{\boldsymbol{M}}}
\newcommand{\bfP}{{\boldsymbol{P}}}
\def\apj{ApJ}
\def\apjl{ApJL}
\def\apjs{ApJS}
\def\aj{AJ}
\def\mnras{MNRAS}
\def\physrep{physrep}
\def\pasj{PASJ}
\def\araa{{Ann.\ Rev.\ Astron.\& Astrophys.\ }}
\def\aap{{\em A.\&A}}
\def\prd{PRD}

\vspace{-0.3cm}
\fb{\bf \underline{Key benefits of joint gravitational/electromagnetic observations:}%\medskip\\
\begin{itemize}
\item Extend the sensitivity of GW detectors, and
improve our determination of signal properties, through association with an EM counterpart

\item Measure cosmological parameters precisely with a novel approach

\item Directly observe and precisely measure properties of the
engine driving violent astrophysical events, such as short-hard GRBs,
supernovae, and black hole mergers
\end{itemize}
}

\vspace{-0.13cm}
\section{Introduction}
\vspace{-0.2cm}

It is widely expected that the coming decade will witness the first
direct detection of gravitational waves (GWs). The ground-based
\href{http://www.ligo.caltech.edu}{LIGO} and
\href{http://www.virgo.infn.it}{Virgo} detectors are being upgraded
to ``advanced'' sensitivity, and are expected to observe a significant
binary merger rate (perhaps dozens per year; e.g., \cite{koppa08}).
The launch of the planned \href{http://lisa.nasa.gov}{\it LISA}
antenna will extend the GW window to low frequencies, opening new vistas
on dynamical processes involving massive ($M \age
10^5\,M_{\odot}$) black holes.

GW events are likely to be accompanied by electromagnetic (EM)
counterparts (e.g., see \cite{2003ApJ...591.1152S,
stubbs08} for review).  Since information carried electromagnetically is
complementary to that carried gravitationally, a great deal can be
learned about an event and its environment if it becomes possible to
measure both forms of radiation in concert (see the ``key benefits'' box above).

Measurements of this kind will mark the dawn of {\bf trans-spectral
astrophysics}, bridging two distinct spectral bands of information.  Our goal in this whitepaper is to summarize some of the added scientific benefits to be found in coordinating observations between
GW sources and their electromagnetic counterparts. In addition, we
suggest some coordinated facility-level approaches and efforts needed
to carry out these observations.

\vspace{-0.3cm}
\section{The Science Enabled by Joint GW/EM observations}
\vspace{-0.2cm}

We now highlight some of the main (anticipated) benefits of joint GW \& EM observations,
ranging from the more secure to the more speculative:

\noindent{\bf Improving Parameter Extraction of GW Events}: With an EM
identification of a transient, many otherwise degenerate GW errors
collapse, greatly increasing the precision with which we determine
source properties from the GWs; luminosity distance measurements are particularly
improved when the source position is known (e.g.,
\cite{hh03,arunetal08}). EM localization obviates the need to marginalize
over source position, greatly reducing the parameter space of search
templates, and correspondingly increasing the observed
signal-to-noise ratio (SNR).  Likewise, an identification drastically cuts down the search range in time, reducing the threshold
signal-to-noise required for confident detection (\cite{kp93}; see \cite{dhhj} for a recent treatment). Conversely, monitoring a time-variable EM counterpart with an origin
in a dynamical bulk flow that is precisely timed and characterized by
the GW signal offers mutual constraints on the source that are not
otherwise available.

\noindent{\bf A New Precision Cosmology Tool}: Binary inspiral sources
are standard ``sirens,'' with a standardization provided only by an
appeal to General Relativity (\cite{schutz86,cf93,hh05}).  Direct
GW measurement of a coalescing binary provides a
distance-ladder-independent measure of the luminosity distance $D_L$
to a source. For massive BH-BH mergers, calculations show that we can
expect {\it LISA} to measure $D_L$ to $<1$--$2\%$ for redshifts $z <
3$, degrading to $\approx 5\%$ for $z \simeq 5$
(\cite{khmf07,lh08}). While $D_L$ may be well measured, the source
redshift cannot be inferred directly: measurements of the (redshifted) binary mass
and the system's redshift are entirely degenerate \citep{cf93}.
An independent measure of the event redshift is therefore required to populate a Hubble diagram and measure cosmological parameters with
GW events. If an EM signature is detected, spectroscopic observations of the event or the galaxy hosting the EM event should be obtainable.   Though the
utility of distance measurement from such events would be limited by
weak gravitational lensing \citep{hh05,whm02}, the complementarity of
this technique to others in this redshift regime means that it should
be subject to very different systematic effects.  Similarly, in the
more local universe, EM events near the edge of the Advanced
LIGO/Virgo volume would yield precision measurements of $H_0$ ($\age$
few \%) {\citep{dhhj}}. {\it We can use inspirals as cosmological
probes only if we associate the gravitational event with an
electromagnetic counterpart. }

\noindent{\bf What is the Nature of Short-Hard Gamma-Ray Bursts?} The
massive star origin of long-soft $\gamma$-ray bursts (LSBs), representing the majority of
GRB events, was definitively established by the observation of
concurrent envelope-stripped supernovae
(see \cite{2006ARA&A..44..507W} for review). While there is now good, albeit indirect,
evidence that short-hard bursts (SHBs) come from an older stellar
population than LSBs (e.g., \cite{bloometal06,nakar07,zhang07}), the origin of
these events is far from established.  Binary mergers (NS-NS or NS-BH)
are commonly believed to be SHB progenitors~\citep{lee07}, but a number of  other origins remain viable. A concurrent GW inspiral event in the
same place and time as a SHB (detected by, e.g., EXIST; \S \ref{sec:exist}) would be the smoking gun for the
origin of these events.  Moreover, the ensemble rates of GRB and
coincident GW detections would establish the distribution of jet
collimation angles in GRBs, crucial for understanding energetics of
the events \citep{2006ApJ...650..281N,2007ApJ...664.1000B}.  Since a
binary's inclination to the line of sight is a direct GW observable
(it sets the ratio of the two GW polarizations), coordinated
observations of these events offer a wealth of insight into the
geometry of jets and subsequent GRB emission; the detailed nature of the event's
GWs may even be able to give insight into the equation of state of neutron star
material {\citep{cutleretal93,vallis00}}.  Finally, the detailed nature
of correlated GW/EM emission is likely to help elucidate the processes
which drive the GRB engine itself.  Especially for NS-BH driven
events, the final merger, disruption and possibly accretion may
radiate in the most sensitive band of GW detectors.  {\it Concurrent
gravitational-wave and electromagnetic observations of short-hard GRBs
will definitively establish whether the engine is a NS-NS or NS-BH
binary merger, or something else.}

\noindent{\bf Constraining Models of Supernovae (SNe) Core-Collapse}:
Stellar core collapse during a SN releases roughly $10^{53}$ ergs of
gravitational binding energy in less than one second. A consensus
understanding of the physics underlying a core-collapse SN is far from established. Nevertheless, it
is apparent that the masses, velocities, and asymmetries involved have
the potential to generate strong gravitational-wave signals
(\cite{2002ApJ...565..430F}; see also \cite{2008arXiv0809.0695O,fryernew}
and references therein).  At a minimum, detection (or non-detection)
of GWs from a SN strongly constrains the rotation of the collapsed
core.  A SN close enough to be a strong GW source is also likely to be
a strong neutrino source.  {\em The triple ``multi-messenger'' view of GWs, neutrinos, and
photons is likely to provide a wealth of knowledge on the SN engine
and perhaps the behavior of matter at nuclear densities.}
%\citep{fryernew}.

\noindent{\bf Viscous Accretion onto Massive Black Holes with Known
Masses and Spins}: GW observations will determine the masses, spin
magnitudes, and orientations of progenitor and remnant BHs with
unprecedented accuracy by any astronomical standard.  Viscous
accretion of material that remains bound to well-characterized remnant
BHs will lead to afterglow EM emission (e.g.,
{\cite{2002ApJ...567L...9A, mp05}}) and provide some of the
best laboratories for the study of AGN and quasar physics.  Monitoring
of time-variable accretion regimes around massive BHs with varied
spins and viewing geometries will directly inform questions about feedback of
massive BHs on their environments.  Binarity may lead to periodic
signatures in the EM output which, coupled with the GWs as a bulk
mass-flow diagnostic, would allow an unprecedented view of this
dynamics.
{\em Coordinated GW/EM observations will provide an unprecedented
laboratory for the study of BH accretion physics.}

\begin{figure}[t]
\label{fig:slice}
\begin{center}
\epsfig{file=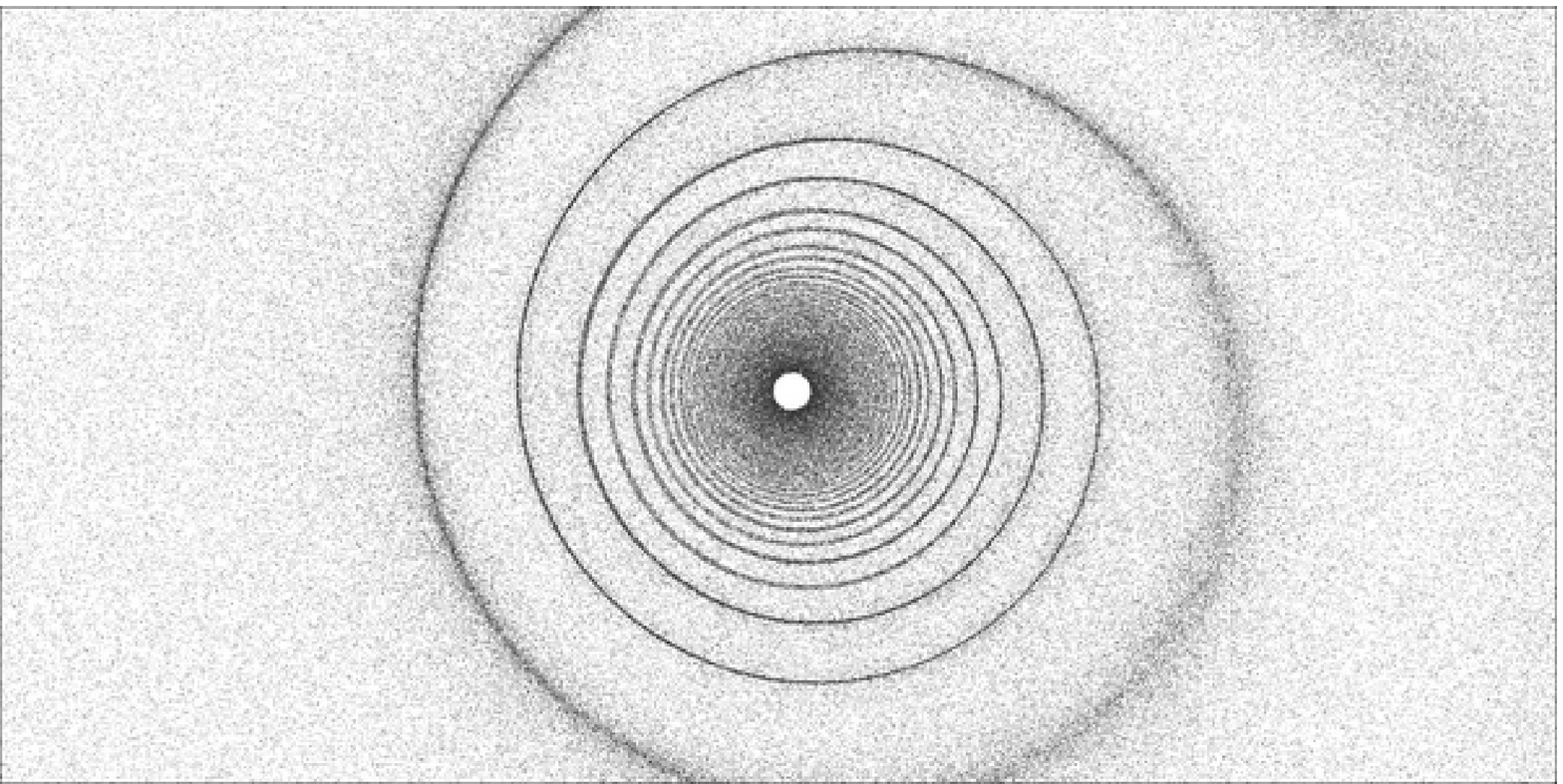, width=8.0cm}\epsfig{file=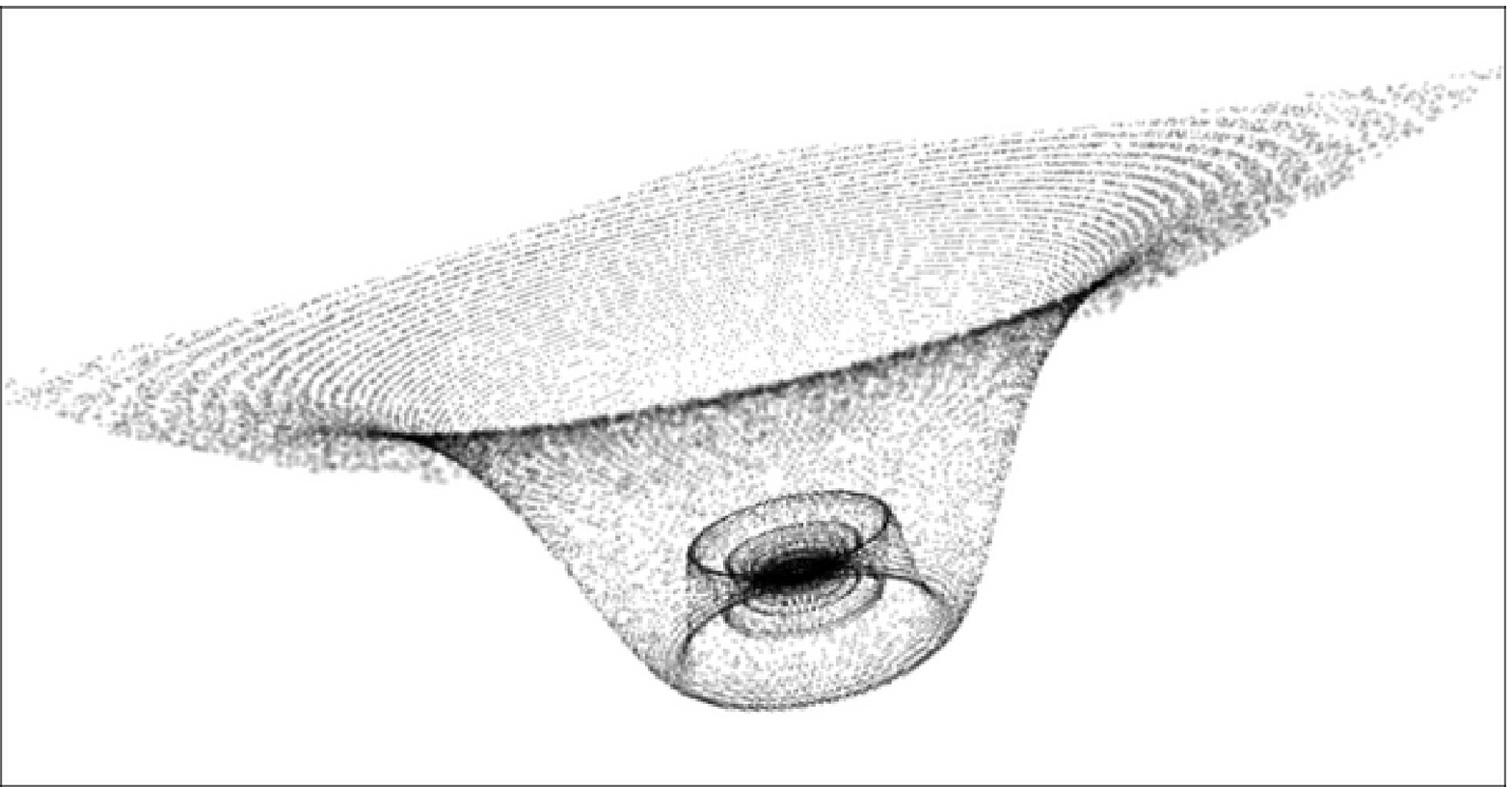, width=8.0cm}
\end{center}
\caption{\it \footnotesize Example scenario for the production of an EM
counterpart from a massive binary BH coalescence.  Typically $\sim$$10^{53}$ erg or greater of GW energy can be channeled as a kinetic
recoil of the merged remnant to be deposited into the surrounding
gaseous environment.  The figures illustrate the induced density waves
in the gas surrounding a $M_1+M_2=10^6~{\rm M_\odot}$ BH binary,
recoiling within the disk plane at a velocity $v_{\rm kick} = 500~{\rm
km~s^{-1}}$.  Strong density enhancements could produce detectable EM
signatures from the perturbed gas.  From \cite{lippaietal08}. }
\end{figure}

\vspace{-0.3cm}
\section{Detecting the Expected EM Signatures}
\vspace{-0.15cm}

The individual nature of coalescing objects will greatly impact what kind of
electromagnetic display might accompany the GW merger
event. EM counterparts can take the form of ``precursors,'' events that
precede the binary coalescence; ``prompt emission,'' events that occur
at (or nearly at) the same time as the coalescence; and
``afterglows,'' emission that follows the GW event:

\noindent{\bf High-Energy Counterparts:} All short-hard GRBs with measured redshifts to date originate from beyond $z=0.1$ (e.g., \cite{2007ApJ...664.1000B}).  A wide-field GRB monitor with today's sensitivities and trigger criteria (or better) should readily detect similar SHBs within the 300 Mpc volume of Advanced LIGO and Virgo.  Coincident
detection in time would virtually guarantee a precise localization of
the GW event.  Combined with GW determination of the source
inclination and the chirp mass, this would open up
a detailed view into the central engine of SHBs that arise from binary
coalescence.  In an exercise that is already becoming commonplace with
the current generation of GW detectors, GRB localizations
in space and time could also be used as an external trigger for more
sensitive searches in GW data streams
\citepeg{2008ApJ...681.1419A}. 

\noindent{\bf Optical/Infrared emission from massive black hole
mergers:} Some GW events are likely to be preceded and/or followed by
detectable optical/infrared emission (e.g., Fig.~1).  Models for such
emission, while not particularly well-developed currently, motivate
deep follow-up searches on GW localizations for EM precursor, prompt
and afterglow characterizations. Optical/infrared emission is readily
tied to candidate host galaxies, and thus redshifts. Transient EM
detections could also be used to trigger localized, higher SNR
searches in {\it LISA} and Advanced LIGO+Virgo data streams
\citep{stubbs08} (see the \cite{sod} WP for specific examples). Deep and wide infrared/optical facilities would be the preferred choices for initial pre-merger searches on larger than
degree scales (Fig.~3), but an increasingly diverse array of
facilities could get involved as localization errors shrink well below
degree scales at late times \citep{lh08} and post-merger. Combining
GW-inferred $D_L$ values with a concordance cosmology yields narrow
redshift slices, from which significantly more focused EM searches can
be carried out. Such information could be used in coordination with
that in \href{http://lsst.org}{LSST}, \href{http://sasir.org}{SASIR} and possibly \href{http://jdem.gsfc.nasa.gov/}{JDEM} photometric redshift catalogs to
weed out unrelated EM transients (see \S 4).

\begin{figure}[tb]
\begin{center}
\begin{minipage}[r]{3.4in}
\centering
\caption{\it \small The first EM signature of a NS-NS \hbox{inspiral}?  Light-curve models
for a $^{56}$Ni-powered ``mini-SN,'' \cite{lp98,kulkarni05,metzgeretal08} compared against optical observations of the transient associated with short-hard GRB 080503. The solid
line indicates a model at $z = 0.03$ with a $^{56}$Ni mass $\sim$$2
\times 10^{-3} M_\odot$, total ejecta mass $\sim$0.4 $M_\odot$, and
outflow velocity $\approx 0.1c$. The dotted line is for a pure
$^{56}$Ni explosion at $z = 0.5$ with mass $\sim$0.3 $M_\odot$ and
velocity $\approx$ 0.2c. From \citet{perleyetal08}.}
\end{minipage}\hspace{0.15in}
\begin{minipage}[l]{2.3in}
\centering
\epsfig{file=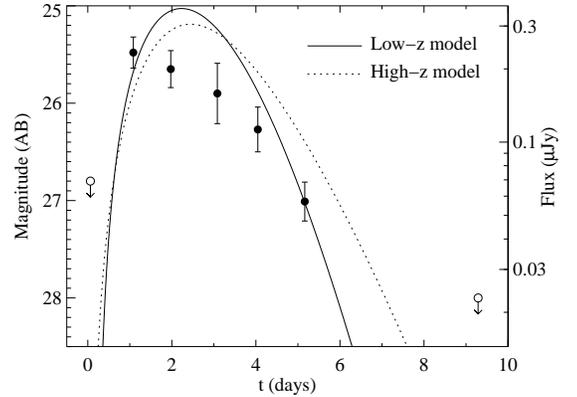,width=3.0in}
\end{minipage}
\end{center}
\label{fig:grb}
\end{figure}

\noindent{\bf Afterglow and SN-like emission from stellar-mass events:}
SHBs are now known to produce faint X-ray, optical, and infrared
afterglows, detectable for $\sim$1 day with current and planned
instrumentation.  NS-NS or NS-BH mergers that do not produce a GRB may
nonetheless produce optical or infrared signatures on some timescale.
Advanced LIGO+Virgo uncertainty regions should be systematically
searched as quickly as possible ($<$ hours) for these afterglows
(which would yield arcsec localizations). Later time observations, on
$\sim$day timescales, would be logistically more feasible but more
uncertain to return a counterpart.  Li \& Paczy\'nski \cite{lp98} predicted thermal
SN-like emission, rising and falling on $\sim$day timescales, from
non-relativistic outflow of the coalescing binary.  Sensitive searches
have thus far failed to find bright Li-Paczy\'nski events
\cite{bloometal06,fynboetal05,perleyetal08}, but events peaking at
$M_V > -16$ mag are not ruled out (see
Fig.~2). \citet{2001MNRAS.322..695H} discussed possible radio and
X-ray precursors to compact object mergers. Likewise, the accretion-induced
collapse (AIC) of a white dwarf to a NS (a possible Type Ia SN channel) may also represent a
powerful source of GWs \cite{2002ApJ...565..430F}
and could produce a SN-like transient lasting $\sim$1 day \cite{2008arXiv0812.3656M}. GW-triggered-only events could be identified electromagnetically through repeated and deep observations on tens of square
degrees for days. To, at minimum, inform expectations, the theory of stellar-mass events
should be better developed this decade. 

\vspace{-0.1cm}
\section{Facilitating the Science}
\vspace{-0.2cm}

Using GWs alone, sources will be localized to rather large fields.
The network of ground-based detectors (the two LIGO sites plus the
European Virgo) can pin down binaries to a field of a few to $\sim$10
square degrees \citep{cavalieretal,blairetal08}; the space-based
detector {\it LISA} will be able to pin down merging black holes to a
field of several $\times$ ten square arcminutes in the best cases, and
a few square degrees more typically \cite{khm08,lh08}.
These localizations demand the ability to monitor $\sim$10 square
degree fields and larger in order to find electromagnetic signatures
accompanying the GWs.
%Once the source is located, however, the sky
%position is determined so well that the distance and source
%inclination can be measured with good precision.
Once the source is localized electromagnetically, the distance and source inclination can be measured through GWs with good precision.
Studies show that
ground-based detectors can measure the distance to a coalescing binary
neutron star system with a fractional accuracy of several percent if
its position is known (Nissanke et al., in prep.); {\it LISA}
can similarly pin down the distance to coalescing supermassive black
holes with percent level accuracy or better \cite{khmf07,lh08}.

\noindent In order to realize the science objectives above, we
advocate the following activities and facilities:

\vspace{-0.15cm}
\subsection{Theory}
\vspace{-0.1cm}

\noindent{\bf Modeling and measurement analysis of binary GWs}.  Given
a binary system, general relativity predicts its future evolution and
emitted GWs with zero free parameters.  The post-Newtonian expansion
of general relativity {\citep{blanchet06}} and numerical relativity
{\citep{pretorius07}} have been extraordinarily successful in modeling
these systems, especially when both members are black holes; our
understanding of the dynamics when one member is a neutron star has
greatly advanced as well (e.g., \cite{st08,etienneetal08,liuetal08}).  We advocate continued
attention to the development of such models and the exploration of
binary parameter space,
%with an eye on the impact that such models
%have on how well a binary's characteristics are pinned down by GW
%measurements.
with a focus on how well a binary's characteristics are pinned down by GW
measurements.
For example, an extension of the analyses described in
\cite{arunetal08} to include the impact that the late merger has on
{\it LISA}'s ability to fix the position of a merging binary may have
%great consequences upon the search field necessary to find the
great consequences, by limiting the search field necessary to find the
event's EM afterglow.

\noindent{\bf Modeling the EM counterpart of massive black hole
mergers}.  The nature of EM emission that is likely to accompany the
merger of two massive black holes is rather poorly understood.  At
this point, we cannot say with great confidence whether the emission
will precede, coincide with, or follow the peak GW emission
(e.g., \cite{mp05,dottietal06,lippaietal08,kl08,sk08}).  Given the
binding energy ($\sim$$10^{60}$ erg) and GW luminosity ($\sim$$10^{57}$
erg/sec) involved in these events, even a modest EM conversion
efficiency is likely to be impressive.  We advocate continued analysis
to understand the likely counterparts to these events in order to more
fruitfully guide searches for their accompanying emission.

\vspace{-0.1cm}
\subsection{Gravitational-wave detectors}
\vspace{-0.1cm}

\noindent{\bf LISA, the Laser Interferometer Gravitational-wave
Antenna}.  This instrument will be a space-based antenna for measuring
GWs in the band from about 0.03 milliHz to 0.1 Hz, corresponding to
sources with orbital periods of seconds to hours.  {\it LISA} is
needed to measure massive black hole coalescences.  As is discussed in
white papers by Prince et al.\ and Madau et al., the rate of such
mergers is expected to be high (several to perhaps dozens per year),
especially for events coming from relatively high redshift ($z \gtrsim
3$).  {\it LISA} will localize these sources to square degrees or
better and measure their distances with a precision of a few percent
or better \citep{khm08,lh08}, with the constituent (redshifted) masses and spins
also well measured \citep{lh06}. Inspirals of white dwarfs into massive black holes
constitute yet another avenue for pre-merger localization of
cosmological GW events with plausible EM counterparts, potentially
resulting in valuable constraints on the local Hubble flow
(\cite{mhk08,sves08}).

\noindent{\bf Advanced LIGO}, the Laser Interferometer
Gravitational-wave Observatory.  LIGO consists of three ground-based
GW detectors at two sites, sensitive to waves in the band from about
10 Hz to a few thousand Hz.  It will be needed to measure waves
coincident with SHBs; in concert with the Advanced Virgo detector (with whom LIGO has joint data cooperation), such events can be
localized to within several square degrees.
The NSF-funded Advanced LIGO project started construction activities in April 2008, and plans to start observation as early as 2014.
Although Advanced LIGO is
not a project that is within the scope of the Astro2010 review, we
include it here to note that much of the science we discuss in this
whitepaper depends on data from this instrument.  As such, we
particularly advocate close coordination between the ground-based GW
data analysis, wide-field optical and infrared imaging, and the high
energy surveys that we discuss next. \vspace{-0.3cm}

\begin{center}
\begin{figure}[tbh]
\begin{minipage}[l]{2.6in}
\centering
\epsfig{file=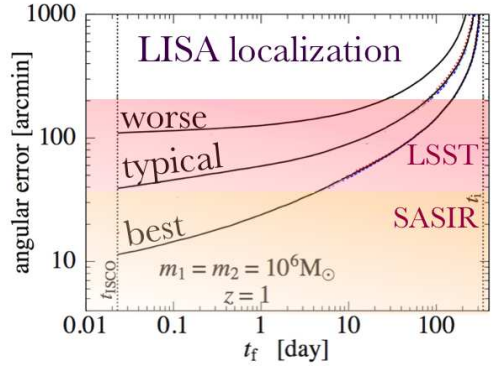,width=2.6in}
\end{minipage}
\hspace{0.5in}
\begin{minipage}[r]{3.3in}
\centering
\caption{\it Pre-merger localization accuracy of a {\it LISA} signal
from two $10^6 M_\odot$ BHs merging at $z=1$, as a function of time before
coalescence, $t_{\rm f}$.  The colored horizontal bands show the {\it
single image} LSST and \href{http://sasir.org}{SASIR} fields of view. The location on the sky
of most events should be easily covered for days to weeks before the
inspiral time {\citep{khmf07,lh08}}. Adapted from
\cite{haimanetal08}.}
\end{minipage}
\label{fig:local}
\end{figure}
\end{center}

\vspace{-1.3cm}

\subsection{Wide-field optical/infrared Imaging}
\vspace{-0.1cm}

\noindent{\bf The Synoptic All-Sky Infrared Imaging (SASIR)} survey is
a pre-phase A concept for a new 6.5m telescope in San Pedro M\'artir
(Mexico) designed to repeatedly observe the infrared sky
simultaneously in 4 bands from 1 -- 2.2 micron. First light is
expected as early as 2016. The 5 sigma survey depth is expected to reach
0.6-1.5 microJy (24.5 -- 23.4 AB mag). A portion of the routine survey
will cover much of the Galactic Plane, improving the possibility of
discovering a heavily obscured Galactic core-collapse SN (a
significant ``blind spot'' in EM observations related to GWs;
\cite{stubbs08}). Roughly $30\%$ of the time has also been budgeted
on fast (and dynamic) cadences in small regions of the sky. Follow-up
of Advanced LIGO+Virgo events, to search for rapidly fading afterglows
and for long-lived transients (e.g. Li-Paczy\'nski mini-SNe), is a
science driver for ToO observations on SASIR
(typical response expected to be $< 2$ min) and field-of-view tiling of {\it LISA}
pre-merger localizations. Combining all-sky IR observations with other
wide-field optical surveys should drastically improve the photometric
redshifts of galaxies hosting $z > 1.4$ BH-BH mergers, allowing for a
more efficient and sensitive search for EM signatures to {\it LISA}
events (where cosmological constraints derived from the observed $D_L$
allow for a narrow search in redshift space). It is important to
emphasize that since dust obscuration might hide optical signatures,
IR observations could prove crucial.

\noindent{\bf The Large Synoptic Survey Telescope (LSST)} is a
$\sim$10 square degree FOV optical imaging survey, reaching single epoch depths of $\sim$0.3$\mu$Jy
in 6 bands, and covering 20,000 square degrees with $\sim$1000 repeat visits. The combination of depth and field of view is a powerful tool for studying EM counterparts of GW events with two major modes of
operation.  First, a quick response (ToO) mode where LSST would slew
to an Advanced LIGO+Virgo GW event location and search for new
transients (likely associated with relatively nearby galaxies). A
similar response could be instituted for {\it LISA} pre-merger
followup. Second, historical data drill down into the regions where GW
events were uncovered and not promptly communicated.  LSST pre-imaging
of GW event sites would provide important clues to uncovering the
nature of the EM signature eventually detected.

\vspace{-0.2cm}
\subsection{High-energy Surveys}
\vspace{-0.1cm}
\label{sec:exist}

\noindent{\bf The Energetic X-ray Imaging Survey Telescope (EXIST)} is
an all-sky hard X-ray imaging facility currently in Pre-Phase A study.
On-board triggering within EXIST's 90 deg $\times$ 70 deg FoV and prompt 
$\gamma$-localization ($<20''$), followed ($< 2$ min) by onboard $0.3 - 2.3\,\mu$m 
imaging (to AB $\approx$ 25 mag) and spectroscopy ($R = 30$ and 3000) ensure maximal 
EM coverage of the $\sim$10 coincident SHB and Advanced LIGO/Virgo events 
per year. Even without a $\gamma$-ray signature (though this is mitgated by EXIST's high sensitivity), the fast slew ToO capability of EXIST would enable immediate deep imaging and spectra with the soft X-ray 
and 1.1m optical-IR telescope, which would locate afterglow emission 
to within $0.1$ arcsec.

\vspace{-0.2cm}
\subsection{Other Facilities}
\vspace{-0.15cm}

Continuous monitoring of {\it LISA} merger locations in the days to hours
preceding merger will require coordinated efforts across the globe
(and in space). Whole-earth monitoring programs across a (probably
heterogenous) network of EM facilities will be required. Such a
paradigm on small-aperture facilities is already producing important results in microlensing research (see WP from Gaudi/Gould et al.). Centralized networks, such as the Las Cumbres
Global Telescope Network (LCOGT), should be well positioned for
monitoring and follow-up activities. Virtual Observatory standards are
already in place to describe and broadcast GW events to an eager EM
follow-up community (e.g., \href{http://en.wikipedia.org/wiki/VOEvent}{\tt VOEvent}).

Although we have focused on optical/infrared facilities for EM
counterpart searches to {\it LISA} events, facilities across the EM spectrum
(particularly at radio wavebands) could prove beneficial and even critical for late-time searches,
especially when GW localization errors can shrink to sub-degree
scales.

%\medskip
\begin{multicols}{3}
\begin{scriptsize}
%\bibliography{gwem}

\begin{thebibliography}{52}
\expandafter\ifx\csname natexlab\endcsname\relax\def\natexlab#1{#1}\fi

\bibitem[{{Abbott} {et~al.}(2008)}]{2008ApJ...681.1419A}
{Abbott}, B., {et~al.} 2008, \apj, 681, 1419, arXiv:0711.1163

\bibitem[{{Armitage} \& {Natarajan}(2002)}]{2002ApJ...567L...9A}
{Armitage}, P.~J., \& {Natarajan}, P. 2002, \apjl, 567, L9, astro-ph/0201318

\bibitem[{{Arun} {et~al.}(2008){Arun}, {Babak}, {Berti}, {Cornish}, {Cutler},
  {Gair}, {Hughes}, {Iyer}, {Lang}, {Mandel}, {Porter}, {Sathyaprakash},
  {Sinha}, {Sintes}, {Trias}, {Van Den Broeck}, \& {Volonteri}}]{arunetal08}
{Arun}, K.~G. {et~al.} 2008, ArXiv e-prints, arXiv:0811.1011

\bibitem[{{Berger} {et~al.}(2007){Berger}, {Fox}, {Price}, {Nakar}, {Gal-Yam},
  {Holz}, {Schmidt}, {Cucchiara}, {Cenko}, {Kulkarni}, {Soderberg}, {Frail},
  {Penprase}, {Rau}, {Ofek}, {Burnell}, {Cameron}, {Cowie}, {Dopita}, {Hook},
  {Peterson}, {Podsiadlowski}, {Roth}, {Rutledge}, {Sheppard}, \&
  {Songaila}}]{2007ApJ...664.1000B}
{Berger}, E. {et~al.} 2007, \apj, 664, 1000, astro-ph/0611128

\bibitem[{{Blair} {et~al.}(2008){Blair}, {Barriga}, {Brooks}, {Charlton},
  {Coward}, {Dumas}, {Fan}, {Galloway}, {Gras}, {Hosken}, {Howell}, {Hughes},
  {Ju}, {McClelland}, {Melatos}, {Miao}, {Munch}, {Scott}, {Slagmolen},
  {Veitch}, {Wen}, {Webb}, {Wolley}, {Yan}, \& {Zhao}}]{blairetal08}
{Blair}, D.~G. {et~al.} 2008, Journal of Physics Conference Series, 122, 012001

\bibitem[{{Blanchet}(2006)}]{blanchet06}
{Blanchet}, L. 2006, Living Reviews in Relativity, 9, 4

\bibitem[{{Bloom} {et~al.}(2006){Bloom}, {Prochaska}, {Pooley}, {Blake},
  {Foley}, {Jha}, {Ramirez-Ruiz}, {Granot}, {Filippenko}, {Sigurdsson},
  {Barth}, {Chen}, {Cooper}, {Falco}, {Gal}, {Gerke}, {Gladders}, {Greene},
  {Hennanwi}, {Ho}, {Hurley}, {Koester}, {Li}, {Lubin}, {Newman}, {Perley},
  {Squires}, \& {Wood-Vasey}}]{bloometal06}
{Bloom}, J.~S. {et~al.} 2006, \apj, 638, 354, astro-ph/0505480

\bibitem[{{Cavalier} {et~al.}(2006){Cavalier}, {Barsuglia}, {Bizouard},
  {Brisson}, {Clapson}, {Davier}, {Hello}, {Kreckelbergh}, {Leroy}, \&
  {Varvella}}]{cavalieretal}
{Cavalier}, F. {et~al.} 2006, \prd, 74, 082004, gr-qc/0609118

\bibitem[{{Chernoff} \& {Finn}(1993)}]{cf93}
{Chernoff}, D.~F., \& {Finn}, L.~S. 1993, \apjl, 411, L5, gr-qc/9304020

\bibitem[{{Cutler} {et~al.}(1993){Cutler}, {Apostolatos}, {Bildsten}, {Finn},
  {Flanagan}, {Kennefick}, {Markovic}, {Ori}, {Poisson}, \&
  {Sussman}}]{cutleretal93}
{Cutler}, C. {et~al.} 1993, Physical Review Letters, 70, 2984, astro-ph/9208005

\bibitem[{{Dalal} {et~al.}(2006){Dalal}, {Holz}, {Hughes}, \& {Jain}}]{dhhj}
{Dalal}, N., {Holz}, D.~E., {Hughes}, S.~A., \& {Jain}, B. 2006, \prd, 74,
  063006, astro-ph/0601275

\bibitem[{{Dotti} {et~al.}(2006){Dotti}, {Salvaterra}, {Sesana}, {Colpi}, \&
  {Haardt}}]{dottietal06}
{Dotti}, M., {Salvaterra}, R., {Sesana}, A., {Colpi}, M., \& {Haardt}, F. 2006,
  \mnras, 372, 869, astro-ph/0605624

\bibitem[{{Etienne} {et~al.}(2008){Etienne}, {Faber}, {Liu}, {Shapiro},
  {Taniguchi}, \& {Baumgarte}}]{etienneetal08}
{Etienne}, Z.~B., {Faber}, J.~A., {Liu}, Y.~T., {Shapiro}, S.~L., {Taniguchi},
  K., \& {Baumgarte}, T.~W. 2008, \prd, 77, 084002, arXiv:0712.2460

\bibitem[{{Fryer} {et~al.}(2002){Fryer}, {Holz}, \&
  {Hughes}}]{2002ApJ...565..430F}
{Fryer}, C.~L., {Holz}, D.~E., \& {Hughes}, S.~A. 2002, \apj, 565, 430,
  astro-ph/0106113

\bibitem[{{Fryer} \& {New}(2003)}]{fryernew}
{Fryer}, C.~L., \& {New}, K.~C.~B. 2003, Living Reviews in Relativity, 6, 2,
  gr-qc/0206041

\bibitem[{{Fynbo} {et~al.}(2005){Fynbo}, {Gorosabel}, {Smette}, {Fruchter},
  {Hjorth}, {Pedersen}, {Levan}, {Burud}, {Sahu}, {Vreeswijk}, {Bergeron},
  {Kouveliotou}, {Tanvir}, {Thorsett}, {Wijers}, {Castro Cer{\'o}n},
  {Castro-Tirado}, {Garnavich}, {Holland}, {Jakobsson}, {M{\o}ller}, {Nugent},
  {Pian}, {Rhoads}, {Thomsen}, {Watson}, \& {Woosley}}]{fynboetal05}
{Fynbo}, J.~P.~U. {et~al.} 2005, \apj, 633, 317, astro-ph/0506101

\bibitem[{{Haiman} {et~al.}(2008){Haiman}, {Kocsis}, {Menou}, {Lippai}, \&
  {Frei}}]{haimanetal08}
{Haiman}, Z., {Kocsis}, B., {Menou}, K., {Lippai}, Z., \& {Frei}, Z. 2008,
  ArXiv e-prints, arXiv:0811.1920

\bibitem[{{Hansen} \& {Lyutikov}(2001)}]{2001MNRAS.322..695H}
{Hansen}, B.~M.~S., \& {Lyutikov}, M. 2001, \mnras, 322, 695, astro-ph/0003218

\bibitem[{{Holz} \& {Hughes}(2005)}]{hh05}
{Holz}, D.~E., \& {Hughes}, S.~A. 2005, \apj, 629, 15, astro-ph/0504616

\bibitem[{{Hughes} \& {Holz}(2003)}]{hh03}
{Hughes}, S.~A., \& {Holz}, D.~E. 2003, Class.\ Quantum Grav., 20, 65,
  astro-ph/0212218

\bibitem[{{Kochanek} \& {Piran}(1993)}]{kp93}
{Kochanek}, C.~S., \& {Piran}, T. 1993, \apjl, 417, L17+, astro-ph/9305015

\bibitem[{{Kocsis} {et~al.}(2008){Kocsis}, {Haiman}, \& {Menou}}]{khm08}
{Kocsis}, B., {Haiman}, Z., \& {Menou}, K. 2008, \apj, 684, 870,
  arXiv:0712.1144

\bibitem[{{Kocsis} {et~al.}(2007){Kocsis}, {Haiman}, {Menou}, \&
  {Frei}}]{khmf07}
{Kocsis}, B., {Haiman}, Z., {Menou}, K., \& {Frei}, Z. 2007, \prd, 76, 022003,
  astro-ph/0701629

\bibitem[{{Kocsis} \& {Loeb}(2008)}]{kl08}
{Kocsis}, B., \& {Loeb}, A. 2008, Physical Review Letters, 101, 041101,
  arXiv:0803.0003

\bibitem[{{Kopparapu} {et~al.}(2008){Kopparapu}, {Hanna}, {Kalogera},
  {O'Shaughnessy}, {Gonz{\'a}lez}, {Brady}, \& {Fairhurst}}]{koppa08}
{Kopparapu}, R.~K., {Hanna}, C., {Kalogera}, V., {O'Shaughnessy}, R.,
  {Gonz{\'a}lez}, G., {Brady}, P.~R., \& {Fairhurst}, S. 2008, \apj, 675, 1459,
  arXiv:0706.1283

\bibitem[{{Kulkarni}(2005)}]{kulkarni05}
{Kulkarni}, S.~R. 2005, ArXiv Astrophysics e-prints, astro-ph/0510256

\bibitem[{{Lang} \& {Hughes}(2006)}]{lh06}
{Lang}, R.~N., \& {Hughes}, S.~A. 2006, \prd, 74, 122001, gr-qc/0608062

\bibitem[{{Lang} \& {Hughes}(2008)}]{lh08}
------. 2008, \apj, 677, 1184, arXiv:0710.3795

\bibitem[{{Lee} \& {Ramirez-Ruiz}(2007)}]{lee07}
{Lee}, W.~H., \& {Ramirez-Ruiz}, E. 2007, New Journal of Physics, 9, 17,
  astro-ph/0701874

\bibitem[{{Li} \& {Paczy{\'n}ski}(1998)}]{lp98}
{Li}, L.-X., \& {Paczy{\'n}ski}, B. 1998, \apjl, 507, L59, astro-ph/9807272

\bibitem[{{Lippai} {et~al.}(2008){Lippai}, {Frei}, \& {Haiman}}]{lippaietal08}
{Lippai}, Z., {Frei}, Z., \& {Haiman}, Z. 2008, \apjl, 676, L5, arXiv:0801.0739

\bibitem[{{Liu} {et~al.}(2008){Liu}, {Shapiro}, {Etienne}, \&
  {Taniguchi}}]{liuetal08}
{Liu}, Y.~T., {Shapiro}, S.~L., {Etienne}, Z.~B., \& {Taniguchi}, K. 2008,
  \prd, 78, 024012, arXiv:0803.4193

\bibitem[{{Menou} {et~al.}(2008){Menou}, {Haiman}, \& {Kocsis}}]{mhk08}
{Menou}, K., {Haiman}, Z., \& {Kocsis}, B. 2008, New Astronomy Review, 51, 884,
  arXiv:0803.3627

\bibitem[{{Metzger} {et~al.}(2008{\natexlab{a}}){Metzger}, {Piro}, \&
  {Quataert}}]{metzgeretal08}
{Metzger}, B.~D., {Piro}, A.~L., \& {Quataert}, E. 2008{\natexlab{a}}, ArXiv
  e-prints, arXiv:0810.2535

\bibitem[{{Metzger} {et~al.}(2008{\natexlab{b}}){Metzger}, {Piro}, \&
  {Quataert}}]{2008arXiv0812.3656M}
------. 2008{\natexlab{b}}, ArXiv e-prints, 0812.3656

\bibitem[{{Milosavljevi{\'c}} \& {Phinney}(2005)}]{mp05}
{Milosavljevi{\'c}}, M., \& {Phinney}, E.~S. 2005, \apjl, 622, L93,
  astro-ph/0410343

\bibitem[{{Nakar}(2007)}]{nakar07}
{Nakar}, E. 2007, \physrep, 442, 166, astro-ph/0701748

\bibitem[{{Nakar} {et~al.}(2006){Nakar}, {Gal-Yam}, \&
  {Fox}}]{2006ApJ...650..281N}
{Nakar}, E., {Gal-Yam}, A., \& {Fox}, D.~B. 2006, \apj, 650, 281,
  astro-ph/0511254

\bibitem[{{Ott}(2008)}]{2008arXiv0809.0695O}
{Ott}, C.~D. 2008, ArXiv e-prints, arXiv:0809.0695

\bibitem[{{Perley} {et~al.}(2008){Perley}, {Metzger}, {Granot}, {Butler},
  {Sakamoto}, {Ramirez-Ruiz}, {Levan}, {Bloom}, {Miller}, {Bunker}, {Chen},
  {Filippenko}, {Gehrels}, {Glazebrook}, {Hall}, {Hurley}, {Kocevski}, {Li},
  {Lopez}, {Norris}, {Piro}, {Poznanski}, {Prochaska}, {Quataert}, \&
  {Tanvir}}]{perleyetal08}
{Perley}, D.~A. {et~al.} 2008, ArXiv e-prints, arXiv:0811.1044

\bibitem[{{Pretorius}(2007)}]{pretorius07}
{Pretorius}, F. 2007, ArXiv e-prints, arXiv:0710.1338

\bibitem[{{Schnittman} \& {Krolik}(2008)}]{sk08}
{Schnittman}, J.~D., \& {Krolik}, J.~H. 2008, \apj, 684, 835, arXiv:0802.3556

\bibitem[{{Schutz}(1986)}]{schutz86}
{Schutz}, B.~F. 1986, Nature, 323, 310

\bibitem[{{Sesana} {et~al.}(2008){Sesana}, {Vecchio}, {Eracleous}, \&
  {Sigurdsson}}]{sves08}
{Sesana}, A., {Vecchio}, A., {Eracleous}, M., \& {Sigurdsson}, S. 2008, \mnras,
  391, 718, arXiv:0806.0624

\bibitem[{{Shibata} \& {Taniguchi}(2008)}]{st08}
{Shibata}, M., \& {Taniguchi}, K. 2008, \prd, 77, 084015, arXiv:0711.1410

\bibitem[{{Soderberg} {et~al.}(2009)}]{sod}
{Soderberg}, A.~M., {et~al.} 2009, Decadal Survey Whitepaper

\bibitem[{{Stubbs}(2008)}]{stubbs08}
{Stubbs}, C.~W. 2008, Class.\ Quantum Grav., 25, 184033, arXiv:0712.2598

\bibitem[{{Sylvestre}(2003)}]{2003ApJ...591.1152S}
{Sylvestre}, J. 2003, \apj, 591, 1152, astro-ph/0303512

\bibitem[{{Vallisneri}(2000)}]{vallis00}
{Vallisneri}, M. 2000, Physical Review Letters, 84, 3519, gr-qc/9912026

\bibitem[{{Wang} {et~al.}(2002){Wang}, {Holz}, \& {Munshi}}]{whm02}
{Wang}, Y., {Holz}, D.~E., \& {Munshi}, D. 2002, \apjl, 572, L15,
  astro-ph/0204169

\bibitem[{{Woosley} \& {Bloom}(2006)}]{2006ARA&A..44..507W}
{Woosley}, S.~E., \& {Bloom}, J.~S. 2006, \araa, 44, 507, astro-ph/0609142

\bibitem[{{Zhang}(2007)}]{zhang07}
{Zhang}, B. 2007, Chinese Journal of Astronomy and Astrophysics, 7, 1,
  astro-ph/0701520

\end{thebibliography}

\end{scriptsize}
\end{multicols}
\end{document}